\documentclass{article}
\usepackage{beamerarticle}

\mode<presentation>{\usetheme{Warsaw}}
\usepackage{pgf,pgfarrows,pgfnodes,pgfautomata,pgfheaps,pgfshade}

\usepackage[english]{babel}
\usepackage[latin1]{inputenc}
\usepackage{xmpmulti}

\setbeamercovered{transparent}

\usepackage{graphics}
\usepackage{bm}
\usepackage{amsmath}
\usepackage{amssymb}
\begin{document}

\title{\bf On the resonance spectra of particle-unstable light nuclei 
with a Sturmian approach that preserves the Pauli principle. }
\author{{L. Canton}$^{(1)}$, K. Amos$^{(2)}$, S. Karataglidis$^{(3)}$,
         G.~Pisent$^{(1)}$, J.~P.~Svenne$^{(4)}$
\\ \ 
%	 
%	 }
%\affiliation{
\\ $^{(1)}$ Istituto  Nazionale  di  Fisica  Nucleare,
   Sez.  di Padova, e\\  Dipartimento di Fisica  dell'Universit\`a, \\
   via Marzolo 8, Padova I-35131, Italia,\\
   $^{(2)}$ School of Physics, The University of Melbourne,\\
   Victoria 3010, Australia\\
   $^{(3)}$ Department of Physics and Electronics,\\
   Rhodes University, Grahamstown 6140, South Africa\\
   $^{(4)}$ Department  of  Physics  and Astronomy,
   University  of Manitoba, \\ and Winnipeg  Institute for   Theoretical
   Physics,\\ Winnipeg, Manitoba, Canada R3T 2N2}
                                                                                
\date{\today}

\mode<presentation>{\frame{\titlepage}}                                                                                
\mode<article>{\maketitle}

\mode<presentation>{
\title{On the resonance spectra of particle-unstable ...}                                                                               
\author{{\alert{\bf VARENNA}}\ \ \ \ \ \ \ \ \ \ \ \ \ \ \ \ \ \ \ L. Canton et al.}

\section*{outlines}

\subsection{Part I :Sturmians}

\frame{
\mode<presentation>{  \nameslide{outline}}
  \frametitle{Sturmians}
  \tableofcontents[pausesections,part=1]
}

\subsection{Part II: Coupled-channel Potential and OPP}

\frame{
\mode<presentation>{  \nameslide{outline}}
  \frametitle{Model Coupled-Channel Potential and OPP}
  \tableofcontents[pausesections,part=2]
}

\subsection{Part III: Applications}

\frame{
\mode<presentation>{  \nameslide{outline}}
  \frametitle{Applications}
  \tableofcontents[pausesections,part=3]
}

}
%%%%%%%%%%%%%%%%%%%%%%%%%%%%%%%%%%%%%%%%%%%%%%%%%%%%%%%%%%%%                                                                                
\mode<article>{

\begin{abstract}
The fundamental ingredients of the MCAS (multi-channel algebraic scattering)
method are discussed. The main feature, namely the application of the 
sturmian theory for nucleon-nucleus scattering, 
allows solution of the scattering problem given
     the phenomenological
ingredients necessary for the description of 
weakly-bound (or particle-unstable) light nuclear systems. 
Currently, to describe these systems, we use a macroscopic, collective 
model. Analyses show that the couplings
to low-energy collective-core excitations are fundamental but they 
are physically meaningful only if the constraints introduced by the 
Pauli principle are taken into account. For this we introduce in the
nucleon-nucleus system the Orthogonalizing Pseudo-Potential formalism, 
extended to collective excitations of the core. The formalism 
leads one to discuss a new concept, Pauli hindrance, which appears to be important 
especially to understand the structure of weakly-bound and unbound systems.
\end{abstract}
%{\LARGE Never underestimate a nuclear model \\
%that reproduces \alert{bound and scattering} \\
%spectra with the same Hamiltonian ! }
%\end{abstract}
%\pacs{24.10-i;25.40.Dn;25.40.Ny;28.20.Cz}
%}
}
%%%%%%%%%%%%%%%%%%%%%%%%%%%%%%%%%%%%%%%%%%%%%%%%%%%%%%%%%%%%%%%%%%%%%%%%%

\mode<presentation>{\part{Sturmians}

\frame{\partpage}
}

\section{Sturmians}
We outline here the use of Sturmian functions in the formulation
of the Coupled-Channel scattering problem.
Sturmians provide the solution of the 
scattering problems by matrix manipulation (hence the word 
``algebraic'' in MCAS).
They are known and used also in atomic and molecular physics, 
chemistry and field theory~\cite{WWS,BMAB}. They provide an
efficient formalism for determination of S-matrices, scattering 
wave functions, bound states, and resonances.
They work well with non-local interactions (such as those non localities
arising from the effects due to Pauli exchanges).
They allow a consistent treatment of Coulomb plus nuclear interactions,
as well as the inclusion in the scattering process of coupled-channel 
dynamics. This occurs, for instance, when strong coupling to 
low-lying excitations of the target nucleus have to be taken into account. 
%\item Algebraic derivation of the Optical potential (DPP)

\frame{
Sturmians ({\em also known as} Weinberg states) represent an {\it alternative} 
way to formulate the Quantum Mechanical problem.

Consider a two-body like Hamiltonian $H=H_o+V$:
then the Schr{\"o}dinger equation is written in the standard 
(time-independent)  way
\begin{equation}
(\alert{E}-H_o)\Psi_{\alert{E}} = V \Psi_{\alert{E}}\, ,
\end{equation}
where \alert{E} is the spectral variable, and $\Psi_{\alert{E}}$ 
is the eigenstate.

Sturmians, instead, are the eigensolutions of:
\begin{equation}
(E-H_o)\Phi_{\alert{i}}(E) = \frac{V}{\alert{\eta_i}(E)} \Phi_{\alert{i}}(E)\, ,
\end{equation}
where E is a parameter. The eigenvalue \alert{$\eta_i$} is the potential scale.
Thus the spectrum consists of all the potential rescalings that give 
solution to that equation, for given energy E, 
and with well-defined boundary conditions.

%Standard boundary conditions of $\Phi_i(E)$:
%\begin{itemize}
%\item[{$E < 0$}] Bound-state like; normalizable.
%\item[{$E > 0$}] Purely outgoing/radiating waves; non-normalizable.
%\end{itemize}
}

\frame{
%For any energy $E$, Sturmians are \alert{$V$}-complete and 
%\alert{$V$}-normalizable:
%\begin{equation}
%V =\sum_i V|\Phi_i\rangle %{1\over\eta_i} 
%\langle \Phi_i|V
%\end{equation}
%\begin{equation}
%%\eta_i
%\delta_{ij}=\langle \Phi_i|V|\Phi_i\rangle
%\end{equation} 
%(\alert{Careful! the normalization is non trivial})

The standard boundary conditions for Sturmians $\Phi_i(E)$ are:
\begin{itemize}
\item[{$E < 0$}] Bound-state like; normalizable.
\item[{$E > 0$}] Purely outgoing/radiating waves; non-normalizable.
\end{itemize}

The spectrum of eigenvalues is purely discrete, and bounded absolutely. 
For short-range (nuclear-type) potentials, the eigenvalues can accumulate
around 0 only.
}

\frame{
Then, the single-channel $S$-matrix can be written as
\begin{equation}
S(E)=\frac{\Pi_i (1-\eta_i(E^{(-)}))}{ \Pi_i (1-\eta_i(E^{(+)}))}
\end{equation}
Alternatively, introducing the factor $\hat\chi_i(E,k)$ in momentum space 
\begin{equation}
\hat\chi_i(E,k) = <k,c| V | \Phi_i(E)> \, ,
\end{equation}
the $S$-matrix can be rewritten also as
\begin{equation}
S(E)= 1 - i \pi k \sum_i \hat\chi_i(E^{(+)};k)\frac{1}{ 1-\eta_i(E^{(+)})}
\hat\chi_i(E^{(+)};k)
\end{equation}
}

\frame{
Most interestingly, the last expression can be generalized to coupled-channel
dynamics: one starts from a coupled-channel Hamiltonian with potential 
$V_{cc'}$, and obtains an $S$-matrix of the form~\cite{Rawit}:
\begin{equation}
S_{cc'}(E)= \delta_{cc'} - i \pi \sqrt{k_c k_c'} 
\sum_i \hat\chi_{ci}(E^{(+)};k_c)\frac{1}{ 1-\eta_i(E^{(+)})} 
\hat\chi_{c'i}(E^{(+)};k_{c'})
\end{equation}

It is remarkable that the following interpretation can be given to the 
last expression: the scattering process initiated in the asymptotic channel 
$c$  is ``captured'' into Sturmians. Subsequently the Sturmian propagates 
freely in the interaction region and finally decays into the outgoing 
channels. This structure naturally reflects the description of 
the scattering process in terms of compound nucleus formation,
and leads to an expression which is rather similar to that obtained
in $R$-matrix formalism. However, in the Sturmian approach, such
structure emerges directly from the Hamiltonian, while
in the R-matrix formalism, the resonant (compound) structure is 
modeled phenomenologically in terms of specific boundary conditions 
given at the surface of an hypothetic  R-space sphere.

\subsection{Resonances and bound states in terms of Sturmians}

Resonant structures as well as bound states can be obtained 
in terms of the properties of Sturmian eigenvalues.
A bound state occurs when one of the eigenvalues moves toward
the right on the real axis, and crosses the value 1 at some negative 
energy. That particular energy value corresponds to the 
bound-state energy. 
         A resonance occurs when the eigenvalue, initially
        progressing along the the real axis, becomes complex before
        reaching the point (1,0). Such occurs for positive energies as
        the scattering threshold (E = 0) must be passed. The energy
        centroid of the resonance 
is the energy corresponding to 
the real part of sturmian eigenvalue matching the value 1. The width of the 
resonance can also be determined geometrically by the patterns 
of the sturmian trajectories, and relates to the imaginary part
of the sturmian eigenvalue at the resonant energy. Such patterns can be 
seen in Fig.~\ref{fig1}. In Fig.~\ref{fig2} one observes the situation in a
realistic case, namely $n$-$^{12}C$ elastic scattering. The eigenvalue
trajectories produce two $3/2^{+}$ resonances that can be clearly seen
in the experimental data as well as in the theoretical calculation.

\frame{
\begin{figure}[h]
\scalebox{0.4}{\includegraphics{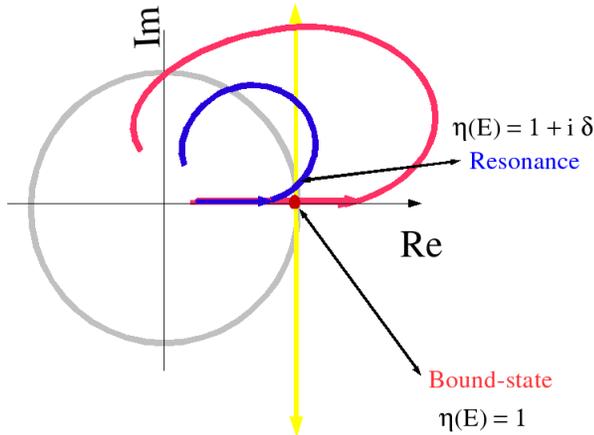}}
\caption{\label{fig1}
How resonances and bound states are found in Sturmian theory.}
%%%%%%%%%%%%%%%%%%%%%%%%%%%%%%%%%%%%%%%%%%%%%%%%%%%%%%%%%%%%%%%%%%%
\end{figure}
}

\frame{
\begin{figure}[ht]

\scalebox{0.35}{\includegraphics{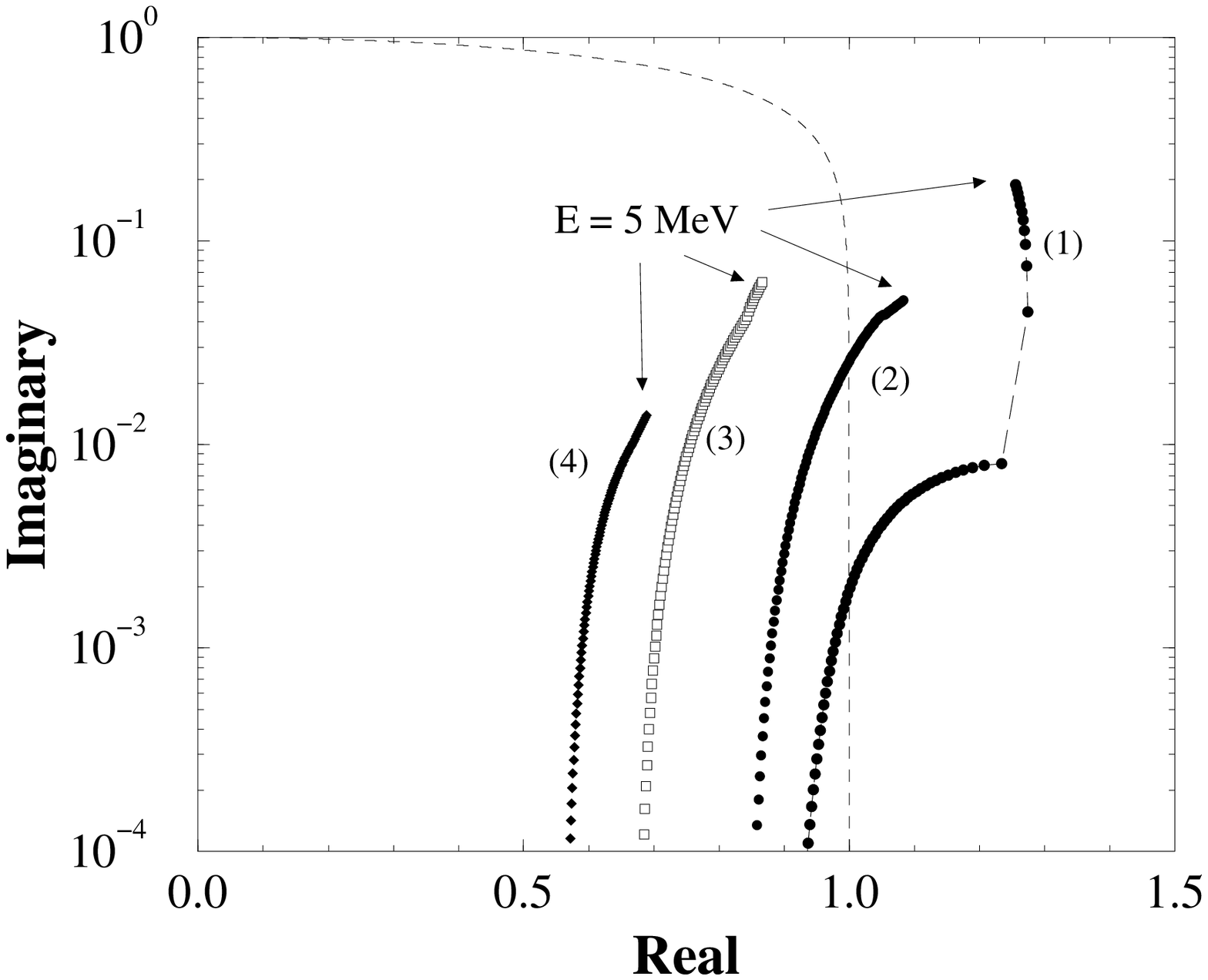}} 
\scalebox{0.58}{\includegraphics{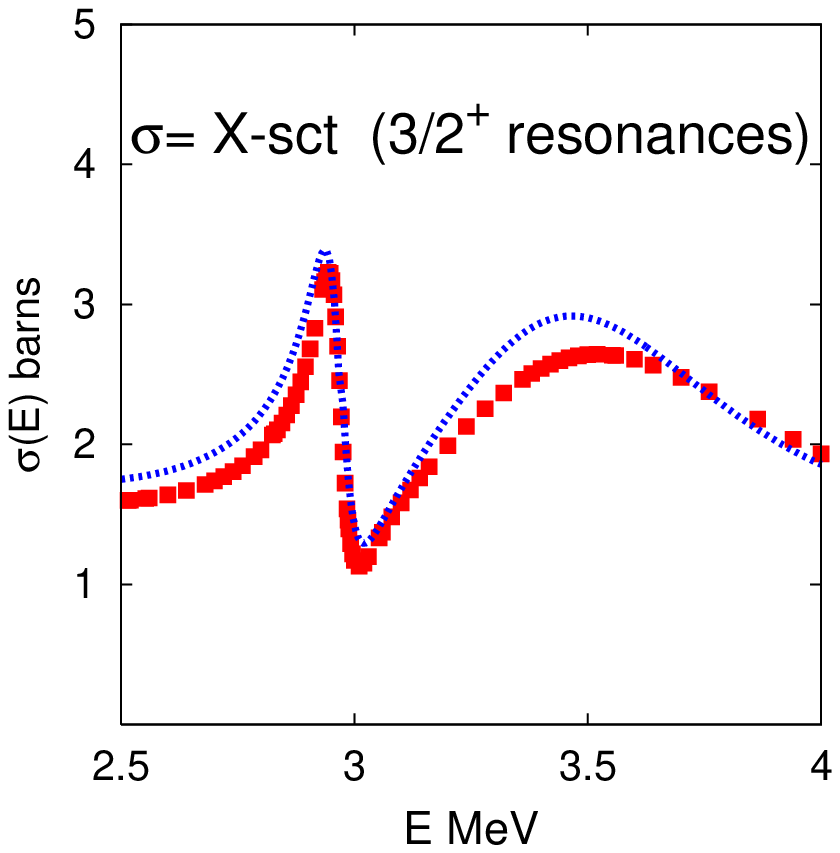}}
\caption{\label{fig2}
Neutron-$^{12}$C in the 3/2$^+$ channel: a realistic 
case. Low-energy resonances in 3/2$^+$ $n$-$^{12}$C 
system. Sturmian patterns (left) and 3/2$^+$ resonant cross-section (right).
The dashed line in the left panel denotes the unit circle.}
\end{figure}
}
%\caption{\label{Fig2}
%}
%\end{figure}

\mode<presentation>{\part{Collective Coupled-Channel potential and OPP}
\frame{\partpage}
}

\section{Model Coupled-channel potential and OPP}

\frame{
To date, we have used a macroscopic potential approach:
a nucleon is scattered by a nucleus (light nuclei with 0$^+$ g.s. are 
considered) and we include couplings to first core excitations 
of collective nature (quadrupole, octupole, etc), since these couplings 
play an important role in the dynamics. The coupled-channel potential
that describes the dynamics (including the couplings to collective-type
target excitations) is an expansion over four operators,
where the first two play a dominant role (central and spin-orbit),
while the remaining two operators (orbit-orbit and spin-spin)
produce small phenomenological corrections. The sum over the various 
operator forms can be written as follows~\cite{Amos}

\begin{equation}
V_{cc'}(r)=\sum_{n=C,\ell s,\ell\ell,sI}V_n
<(\ell s)jI;J^\pi|{\cal O}_n f_n(r,R,\theta_{\bf r, R})|(\ell' s)j'I';J^\pi>
\nonumber
\end{equation}

For all operators, the functional forms are expanded to second order 
in the core-deformation parameter. (For simplicity, we discuss here
the case of a single quadrupole deformation effect:
$R=R_0(1+\beta_2P_2(\theta)$.)
\begin{eqnarray*}
f_n(r,R,\theta)=f^{(0)}_n(r)-\beta_2R_0P_2(\theta)\frac{d}{dr}f^{(0)}_n(r)
\\ \ \ \ \ \ \ \ \ \ \   + \frac{\beta_2^2R_0^2}{2\sqrt{\pi}}
\left( P_0-\frac{2\sqrt{5}}{7}P_2(\theta)+\frac{2}{7}P_4(\theta)\right)
\frac{d^2}{dr^2}f^{(0)}_n(r)
\end{eqnarray*}
}

\frame{
The radial forms $f^{(0)}(r)$ are spherically-symmetric functions.

For the central, orbit-orbit, and spin-spin terms we consider a standard
Wood-Saxon form:
\begin{equation}
f^{(0)}_n(r)=[1+\exp^{\frac{{\bf r-R}}{a}}]^{-1} 
\ \ \ \ \ for\ \ (n= C, ll, sI)
\end{equation}

For the spin-orbit term we consider ${\cal O}_{ls} = {\mathbf l\cdot s}$ 
and not the full Thomas term, with the following radial form:
\begin{equation}
f^{(0)}_{LS}(r)=\frac{1}{r}\frac{d}{dr}
[1+\exp^{\frac{{\bf r-R}}{a}}]^{-1} 
\ \ \ \ \ for\ \ (ls)
\end{equation}
}

%\frame{
%\begin{figure}[b]

%\centerline{table of parameters}
%\scalebox{0.8}{\includegraphics{C12-levels.eps}}
%\caption{\label{Fig3}
%The core spectrum used as inut.}
%\end{figure}
%}

\subsection{First application: n-$^{12}$C}
\frame{
In our first application, we considered the scattering of
neutrons off $^{12}$C coupling the ground state of the target
to  the first two low-lying 
excitations $2_1^+$ (4.43~MeV) and $0_2^+$ 
(7.63~MeV), and searched parameters to obtain a description
of the resonant spectra and scattering cross 
sections~\cite{Amos,Canton,Pisent}.
However, many deeply-bound spurious states occurred
in the bound spectrum. It was not possible to obtain a consistent
description of both bound structure and scattering data with 
the same Coupled-Channel Hamiltonian. These spurious
deeply-bound states originate from the violation
of the Pauli principle. The phase-space corresponding to 
target nucleons in fully occupied shells has to be inhibited 
to the incoming nucleon. In the original CC model such condition 
was missing.

Various methods have been suggested to remove the deeply-bound
forbidden states. A recent article~\cite{Brink} on the subject
contains an historical review on the various approaches and their
connections.
In phenomenological macroscopic-type calculations,
such Pauli condition has been implemented first by 
introducing the Orthogonality
Condition Model~\cite{Saito}. Alternatively, the orthogonality condition
can be introduced directly in the Hamiltonian by addition of a new term
in the potential, the highly non-local Orthogonalizing 
Pseudo-Potential~\cite{Kukulin}. With the advent of super-symmetric
quantum mechanics\cite{Witten}, it was possible to define super-symmetric
transformations\cite{Baye} that produce new local (and highly singular) 
potentials which also generate spectra free of spurious states.

In our MCAS approach, we use the technique of Orthogonalizing 
Pseudo Potentials~\cite{Canton}, which eliminates the deep bound states
adding a new term in the nuclear potential.
}

\frame{
The ``complete'' nuclear potential for the $n+{}^{12}$C 
case we considered, has the form
(in partial-wave decomposition)
\begin{eqnarray*}
{\cal V}_{cc'}(r,r')=
V_{cc'}(r)\delta{(r-r')} +
 \delta_{cc'} 
\lambda_{c} 
A_{c}(r)
A_{c}(r') (\delta_{c=s\frac{1}{2}})
+ \delta_{cc'}
\lambda_{c} 
A_{c}(r)
A_{c}(r')  (\delta_{c=p\frac{3}{2}}) \, .
\end{eqnarray*}
 $A_c(r)$ are the 
\alert{Pauli-forbidden} deep (CC-uncoupled) bound states.
The $\lambda$ parameter eliminates
the deeply-bound spurious states, since  
{a state in the OPP approach is }
{\alert{\em forbidden} in the  limit $\lambda\rightarrow +\infty$},
while it is
{{\em allowed}  when $\lambda\rightarrow 0$}
}

\frame{Thus, in ${}^{13}$C, we considered two shells (the $0s_{\frac{1}{2}}$
     and $0p_{\frac{3}{2}}$ shells) to be Pauli blocked.

%\begin{figure}[b]

%\centerline{Bound-state levels and total cross-sect}
%\centerline{\scalebox{0.7}{\includegraphics{C13-levels-OPP-fig.eps}}
%\scalebox{0.3}{\includegraphics{n-c12-sect.eps}}}
%\caption{\label{Fig5}

%\end{figure}

%\centerline{Total Cross-section at low energy}
}

\subsection{The {${\lambda}$}-dependence}

\setbeamercovered{covered}
\begin{frame}
%\centerline{The $\lambda$ dependence 
% \alert{\only<1>{100}
%\only<2>{200}\only<3>{300}\only<4>{500}\only<5>{1000}\only<6>{5000}}}
%\animate<2-4>
%\scalebox{0.6}{\multiinclude[format=eps]{n-C12-sig-log-lambda}}
\begin{figure}

\scalebox{0.5}{
\includegraphics{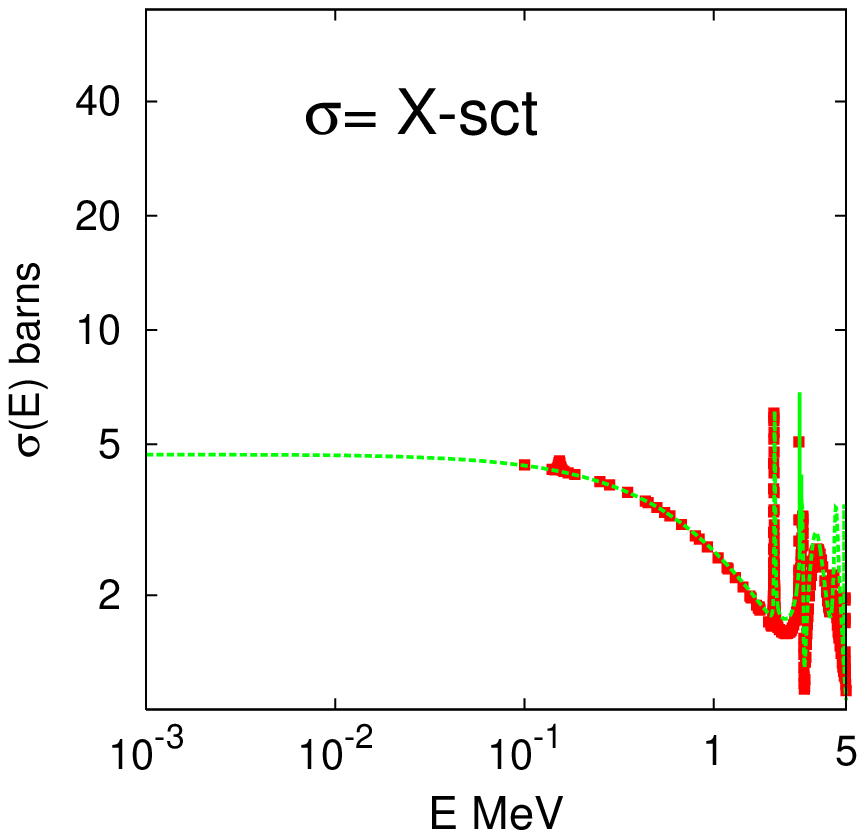}
\includegraphics{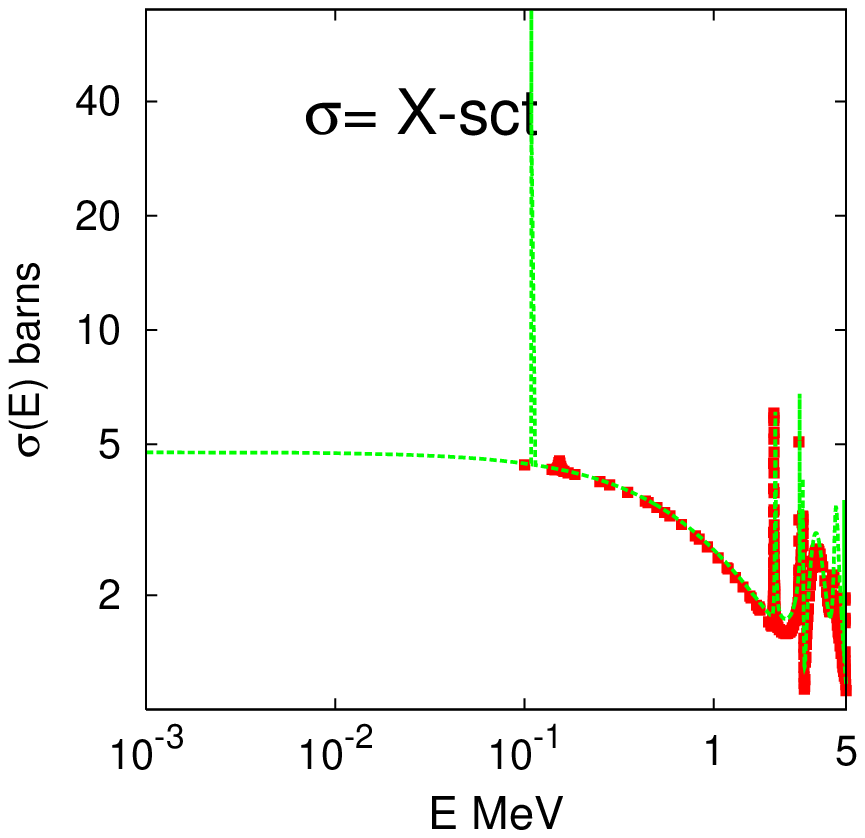}}

\scalebox{0.5}{
\includegraphics{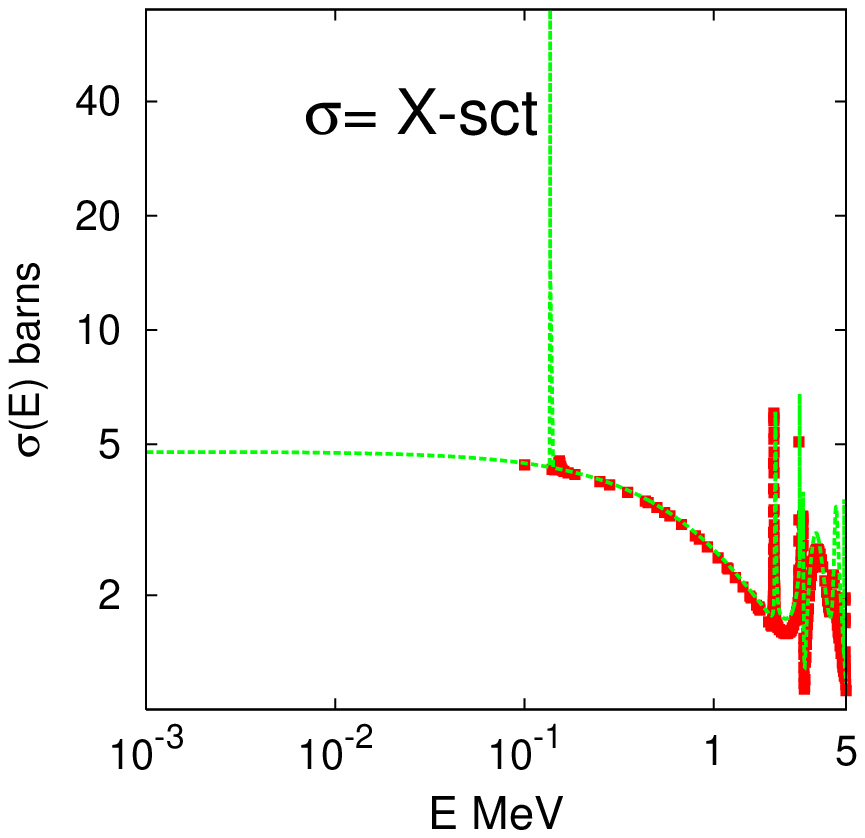} 
\includegraphics{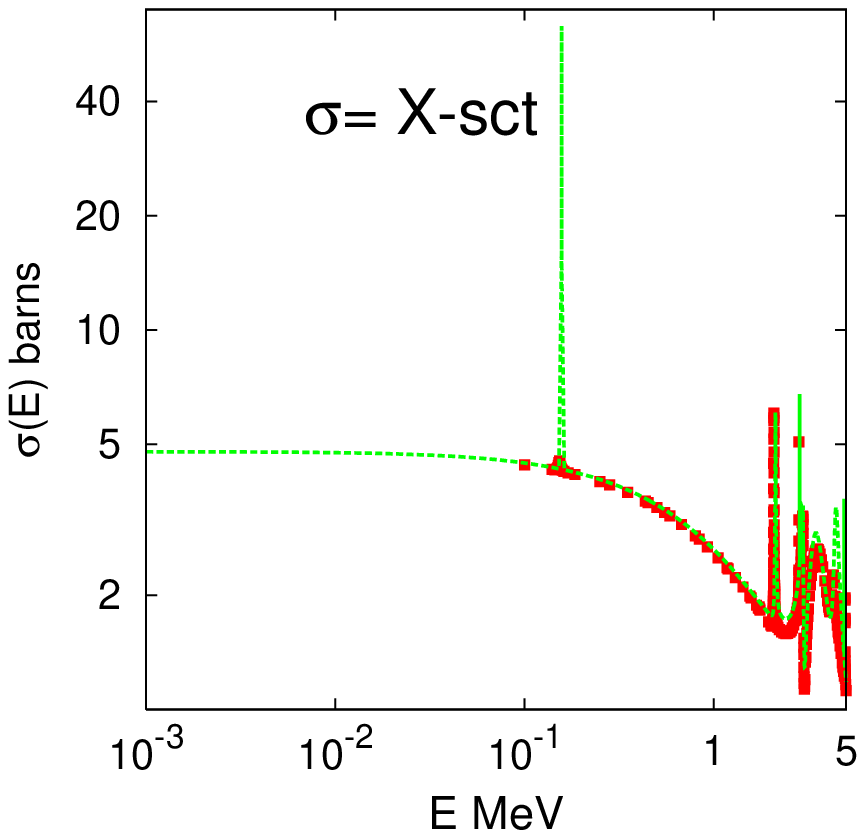}}
\caption{\label{Fig3}
The $\lambda$ dependence (of the elastic cross section
for n-$^{12}C$ scattering) with values 
100, 300, 500, and 1000 Mev (top-left, top-right, bottom-left 
and bottom-right, respectively)}
\end{figure}
%\multiinclude{n-C12-sig-log-lambda}

One important aspect in the OPP method it to assess
the behavior with respect to the $\lambda$ parameters.
Originally, $\lambda$ was set to 100 MeV since this was sufficient
to remove all the spurious states, but later it was
found that some resonances were still sensitive to higher values of 
$\lambda$, corresponding to a stronger orthogonality condition, 
and the energy centroids of those selected resonances improved 
by using larger values of $\lambda$. In particular, 
a narrow 5/2$^-$ resonance, which lies very close to threshold, 
illustrates this well. When $\lambda = 100$ MeV, it is almost a zero-energy 
bound state. For higher values of $\lambda$ that resonance converges
to a position
in agreement with the peak observed in the evaluated data 
around 0.1 MeV. Also the energy centroid of the 1/2$^-$ state around 2.9
MeV stabilizes  with higher $\lambda$ values. These effects indicate
of the need of strong orthogonality conditions in the system.
For further details, including the parameters set for the potential,
we refer the reader to Refs.~\cite{Amos,Canton}.

\end{frame}

\section{Applications of MCAS theory}

\subsection{Analyzing powers of nucleons off $^{12}C$}

\frame{

%\centerline{Analyzing powers of nucleons off $^{12}C$}

Analyses of n-$^{12}$C system were published in
Refs.~\cite{Amos,Canton,Pisent}.
Note that all the parameters, including the spin-orbit term, have 
been fixed on the known spectrum of $^{13}C$. Shortly after our analyses
have been completed, 
new spin-polarization data measured at TUNL were published~\cite{TUNL}.
Without any parameter adjustments or tuning, we could reproduce 
those low-energy data, as shown in Fig.~\ref{Fig4}. Thus, this was a test of our spin-orbit parametrization
and a validation of the spin-structure of our interaction~\cite{Svenne}.

%\frame{
%Scattering amplitudes:
%\begin{eqnarray*}
%A(\theta)=f_C(\theta)+ \frac{1}{2ik} \sum_\ell e^{2i\sigma_\ell}
%\left[ (\ell+1)S^>_\ell+\ell S^<_\ell - (2\ell+1)\right] P_L(\cos{\theta}) 
%\\
%B(\theta)={\frac{1}{2ik}} \sum_\ell e^{2i\sigma_\ell}
%\left[ S^>_\ell - S^<_\ell \right] P^1_L(\cos{\theta}) 
%\end{eqnarray*}
%
%Observables:
%\begin{eqnarray*}
%\frac{d\sigma}{d\Omega}={|A|^2+|B|^2}
%\\
%P=A_y(\theta)= \frac{2\Im A B^*}{|A|^2+|B|^2} 
%\end{eqnarray*}
%}

\setbeamercovered{transparent}

\frame{

\begin{figure}
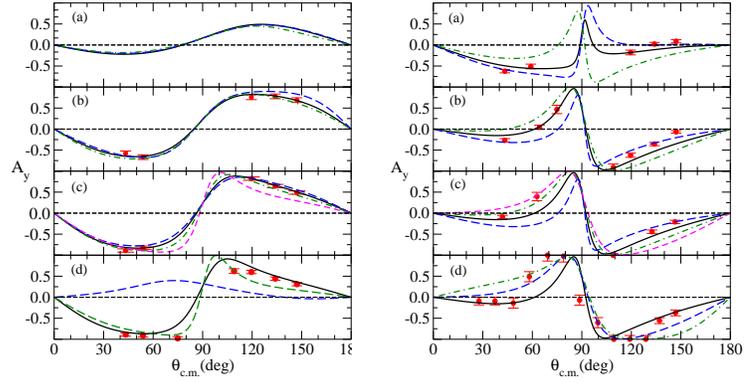


\begin{center}
\begin{columns}
\begin{column}[T]{5.0cm}
{{{\includegraphics[height=5.0cm]{Fig3-appl1.eps}}}}
\end{column}
\begin{column}[T]{5.0cm}
{{{\includegraphics[height=5.0cm]{Fig4-appl1.eps}}}}
\end{column}
\end{columns}
\end{center}
\caption{\label{Fig4}
Neutron analyzing power $A_y$ at various lab energies in the 
MeV ranges (a) 1.9, (b) 2.2, (c) 2.8, (d) 3.2 for the left panel,
and (a) 3.41, (b) 3.62, (c) 3.78, (d) 3.92 for the right panel.}
\end{figure}
}

\subsection{Low-lying unbound states of $^{15}C$ and $^{15}F$}

Our second analysis concerned an unbound nucleus, $^{15}F$, whose
properties were studied in connection with its weakly-bound 
mirror partner $^{15}$C. Our analysis was triggered by recent low-energy 
experimental data on the $^{14}$O-proton system, which we could analyze by
using inverse kinematics.
We included in the model low-lying excitations of $^{14}$O/$^{14}$C, 
in a macroscopic coupled-channel model with parameters given in 
Table ~\ref{newParams}.

\begin{table}
\begin{equation}
\nonumber
\begin{array}{|lll|}
\hline
 V_0^{(\pm)} = -45.0\ {\rm MeV}  & V_{ll}^{(\pm)} = 0.42\ {\rm MeV}  & 
 V_{ls}^{(\pm)} =  7.0\ {\rm MeV} \\
\hline
 R_0 = 3.1\ {\rm fm} & a_0 = 0.65 \ {\rm fm} &   \beta_2 = -0.50\ \\
\hline
\end{array}
\end{equation} 
\caption{\label{newParams}{Coupled-channel parameters} 
for n-$^{14}$C/p-$^{14}$O  }
\end{table}

{\begin{figure}

{\includegraphics[height=5.5cm]{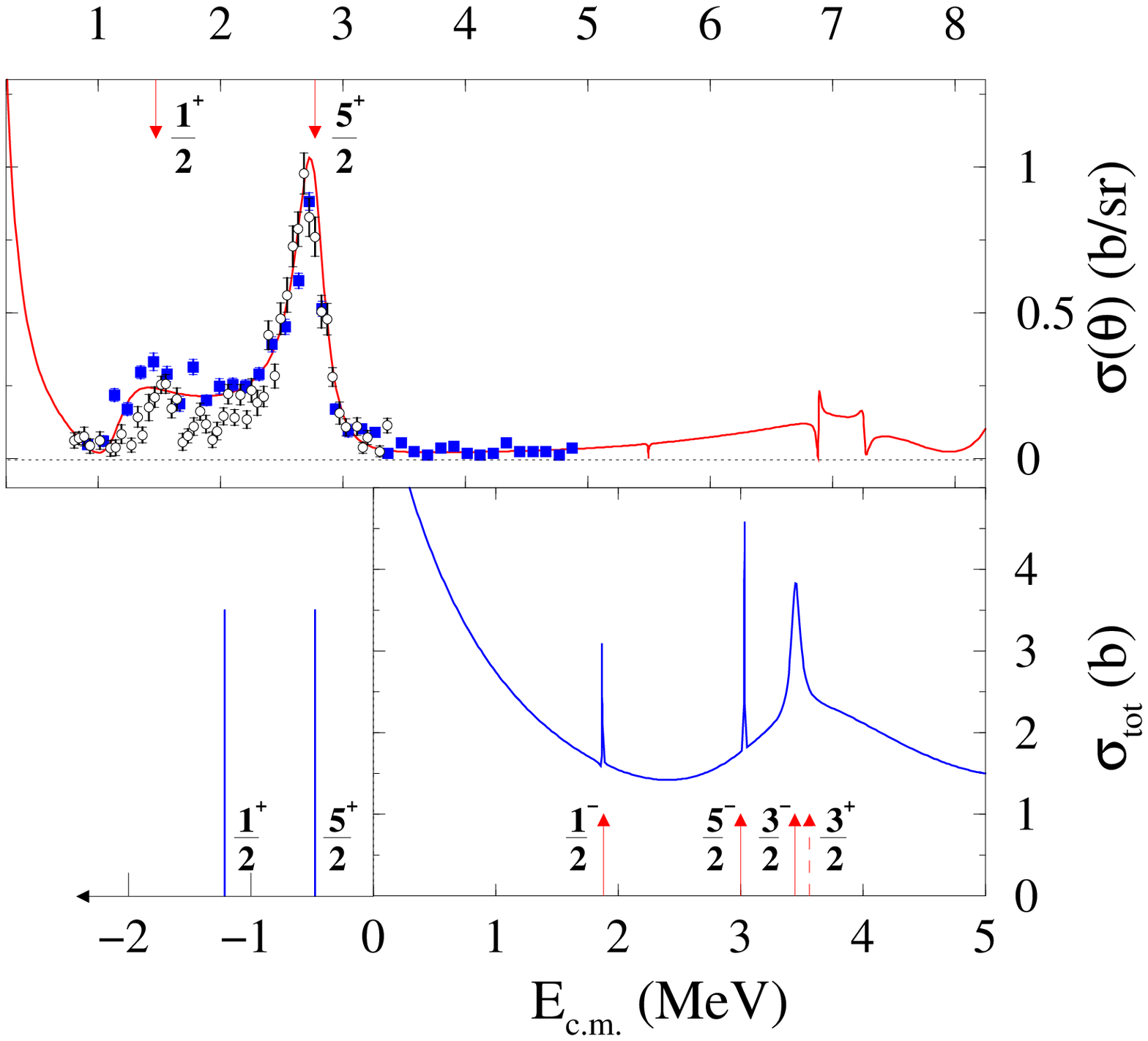}} 
{\includegraphics[height=5.5cm]{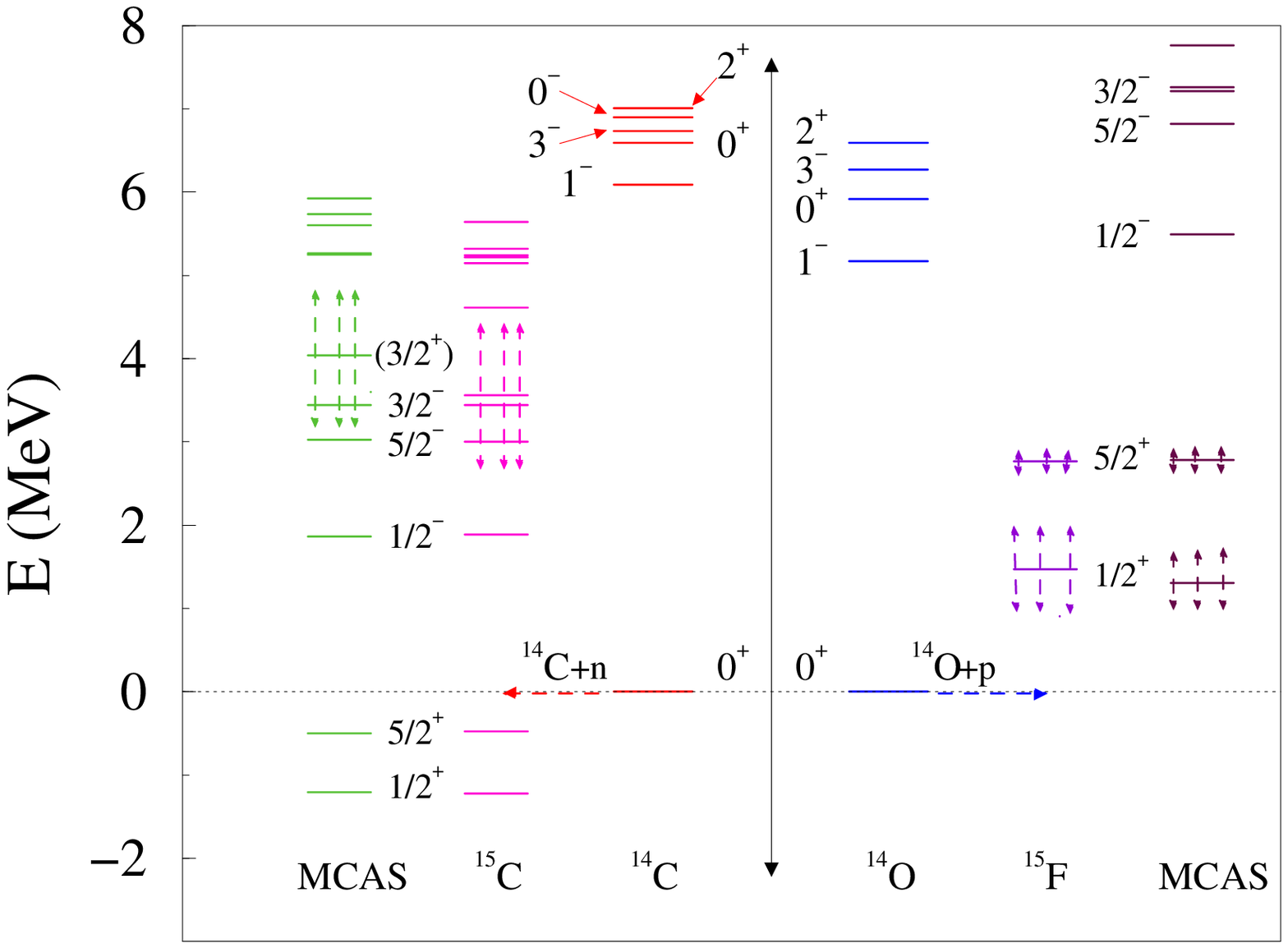}}
\caption{\label{Fig5}
{\em Left:}
$^{15}F$  (top, $p-^{14}O$ differential cross-section\@ $\theta_{cm}=180^o$) 
\& $^{15}C$ (bottom, the bound spectrum and the elastic 
scattering cross-section). {\em Right}: The resonant spectra for $^{15}$C and
$^{15}$F}
\end{figure}
%\end{beamercolorbox}
}

Our results are shown in Fig.~\ref{Fig5}. In the evaluations,
 we used known properties
of $^{15}$C to predict new states in $^{15}$F, in particular
three narrow resonances of negative parity 
(1/2$^-$, 5/2$^-$ and 3/2$^-$ in the range 5-8 MeV). 
To obtain these results, a new concept had to be introduced~\cite{Canton06}:
it is the concept of Pauli-hindered states. Up to now
we have considered states that are Pauli allowed ($\lambda\simeq 0$)
or Pauli forbidden ($\lambda\ge 1 GeV$). Now we introduce 
a state that is neither prohibited 
nor allowed but simply suppressed, or {\em hindered}, by the Pauli principle.
We expect that this situation can apply in weakly-bound/unbound 
light systems where the formation of shells may become critical.
These hindered states can be conveniently described in the OPP
scheme with values of $\lambda$ of a few MeV.  Thus the scheme
of phase-space that is accessible to the proton interacting
with the $^{14}$O core is:

{\small
{\sc Pauli Forbidden ($\lambda\simeq 1 GeV$):}
\begin{eqnarray*}
\alert{0s_{1/2} + 0^+_1} & \alert{0s_{1/2} + 0^+_2} & \alert{0s_{1/2} +
2^+_1}  \\
\alert{0p_{3/2} + 0^+_1} & \alert{0p_{3/2} + 0^+_2} & \alert{1p_{3/2} +
2^+_1} \\
\alert{0p_{1/2} + 0^+_1} & - & -  \\
\end{eqnarray*}
\vskip-0.5truecm
{\sc {Pauli Hindered} ($\lambda\simeq 1 - 50 MeV$):}
\begin{eqnarray*}
\ \ \ \ -\ \ \ \  & \alert{0p_{1/2} + 0^+_2} & \alert{0p_{1/2} + 2^+_1}  \\
\end{eqnarray*} 
\vskip-0.5truecm
{\sc Pauli Allowed ($\lambda = 0 MeV$):}
\begin{eqnarray*}
1s_{1/2} + 0^+_1 & {1s_{1/2} + 0^+_2} & {1s_{1/2} + 2^+_1}  \\
0d_{5/2} + 0^+_1 & {0d_{5/2} + 0^+_2} & {0d_{5/2} + 2^+_1}  \\
0d_{3/2} + 0^+_1 & {0d_{3/2} + 0^+_2} & {0d_{3/2} + 2^+_1}  \\
\cdots & \cdots & \cdots \\
\end{eqnarray*}
}

The need to consider intermediate situations between 
Pauli blocking and Pauli allowance has been registered before in the
literature, mostly in connection with RGM approaches~\cite{Schmidt,Langanke}.

\subsection{P-shells in mass=7 nuclei}

As a third application, we consider the spectral structure the 
mass=7 isobars: $^7$He, $^7$Li, $^7$Be, and $^7$B. We describe 
these systems in terms of a single nucleon-nucleus interaction, explicitly
including the low-lying core excitations of the mass-6 sub-system.
Thus, we use the MCAS approach to determine the bound and resonant
(above nucleon emission) spectra starting from a collective 
coupled-channel interaction model coupling the $g.s.$ of $^6$He
to the first and second $2^+$ excitations.
This description of the mass-7 systems is in many ways alternative
to those based on cluster models ($^7$Li as a $^3H$+$\alpha$ dicluster, 
etc.) or to those based on more microscopic models, either no-core 
shell-model or Green's function Montecarlo 
calculations. A comparison amongst different descriptions (dicluster model,
no-core shell model, and collective-coupling model)
for mass-7 nuclei is given in Ref.~\cite{Canton06a}. 
In the collective-coupling model,
the four A=7 nuclei are described in terms of the system
{\alert{nucleon + mass-6-type nucleus}} with a potential whose parameters
are listed in Table~\ref{OMparams}:
\begin{frame}
\begin{eqnarray*}
^7Li\leftrightarrow p & + & {\alert{^6He(0_1^+[g.s.]; 2_1^+[1.78MeV];
2_2^+[5.6MeV])}}  \\
^7He\leftrightarrow n & + & {\alert{^6He(0_1^+[g.s.]; 2_1^+[1.78MeV];
2_2^+[5.6MeV])}}  \\
^7Be\leftrightarrow n & + & {\alert{^6Be(0_1^+[g.s.]; 2_1^+[1.70MeV];
2_2^+{\color[rgb]{0.2,0.2,0.2}[5.6MeV]})}}  
  \\
^7B\leftrightarrow p & + & {\alert{^6Be(0_1^+[g.s.]; 2_1^+[1.70MeV];
2_2^+{\color[rgb]{0.2,0.2,0.2}[5.6MeV]})}} 
\end{eqnarray*}
\end{frame}

  \begin{table}
  \begin{center}
   \begin{tabular}{|cccc|}
   \hline
   $V_0=$ $-$36.817 & $V_{\ell \ell}=$ $-$1.2346 & $V_{\ell s}=$ 14.9618 & 
   $V_{Is}=$ 0.8511 \\
   \hline
   $R_0=$ 2.8 fm; & $a=$ 0.88917 fm;  & $\beta_2=$ 0.7298 & \\
   \hline
   \end{tabular}
  \caption{\label{OMparams} 
   Parameter values of the nucleon-nucleus(mass-6) potential.}
    \end{center}   
   \end{table} 

\begin{table}

\begin{columns}
\begin{column}[T]{5.0cm}
%\begin{center}
\scalebox{0.6}{\begin{tabular}{c|cc|cc}
  $J^\pi$  & \multicolumn{2}{c}{${}^7$Li} & \multicolumn{2}{c}{${}^7$Be}\\ 
 \hline                                                  
 & Exp. & Theory & Exp. & Theory \\
  \hline
$\frac{3}{2}^-$   & $-$9.975         & $-$9.975      &    $-$10.676          & $-$11.046       \\
$\frac{1}{2}^-$   & $-$9.497         & $-$9.497      &    $-$10.246          & $-$10.680       \\
$\frac{7}{2}^-$   & $-$5.323 [0.069] & $-$5.323      &    $-$6.106 [0.175]   & $-$6.409        \\
$\frac{5}{2}^-$   & $-$3.371 [0.918] & $-$3.371      &    $-$3.946 [1.2]     & $-$4.497        \\
$\frac{5}{2}^-$   & $-$2.251 [0.08]  & $-$0.321      &    $-$3.466 [0.4]     & $-$1.597        \\
$\frac{3}{2}^-$   & $-$1.225 [4.712] & $-$2.244      &          $--$         &    $--$         \\
$\frac{1}{2}^-$   & $-$0.885 [2.752] & $-$0.885      &                       & $-$2.116        \\
$\frac{7}{2}^-$   & $-$0.405 [0.437] & $-$0.405      &    $-$1.406 [?]       & $-$1.704        \\
$\frac{3}{2}^-$   &      $--$        &    $--$       &    $-$0.776 [1.8]     & $-$3.346        \\
$\frac{3}{2}^-$   &    1.265 (0.26)  & 0.704 (0.056) &     0.334 (0.32)       & $-$0.539       \\
$\frac{1}{2}^-$   &                  & 1.796 (1.57) &                     &  0.727 (0.699)     \\
$\frac{3}{2}^-$   &  3.7 (0.8) ?$^a$ & 2.981 (0.99)&                 & 1.995 (0.231)   \\
$\frac{5}{2}^-$   &  4.7 (0.7) ?$^a$ & 3.046 (0.75)&                  & 2.009 (0.203)   \\
$\frac{5}{2}^-$   &                  & 5.964 (0.23)  &                       & 4.904 (0.150)   \\
$\frac{7}{2}^-$   &                  & 6.76  (2.24)  &      6.5 (6.5) ?$^a$  & 5.78  (1.65)   \\
\hline
\end{tabular}}				                          
% \scalebox{0.4}{$^a$For these states spin and parity are unknown}
 %\begin{center}					                          
 %\end{center}					                          
% \scalebox{0.5}{$^b$Spin-parity of this state has been assigned as $\frac{1}{2}^-$}
%\end{center}					                          
%\begin{center}
\end{column}
\begin{column}[T]{5.0cm}
\scalebox{0.6}{\begin{tabular}{c|cc|cc}
$J^\pi$  & \multicolumn{2}{c}{${}^7$He} & \multicolumn{2}{c}{${}^7$B} \\
\hline                                                  
 & Exp. & Theory & Exp. & Theory\\
\hline
$\frac{3}{2}^-$   & 0.445 (0.15)     & 0.43 (0.1)  &    2.21 (1.4)    & 2.10 (0.19)       \\
$\frac{7}{2}^-$   &  $--$            & 1.70 (0.03) &                & 3.01 (0.11)     \\
$\frac{1}{2}^-$   &  1.0 (0.75) ?$^a$ & 2.79 (4.1)  &             & 5.40 (7.2)      \\
$\frac{5}{2}^-$   & 3.35 (1.99)     & 3.55 (0.2)  &                & 5.35 (0.34)      \\
$\frac{3}{2}^-$   & 6.24 (4.0) ?$^a$  & 6.24 (1.9)  &                     &                  \\
\hline
\end{tabular}}				                          
%\scalebox{0.5}{$^a$ Observed very recently and interpreted as a $\frac{1}{2}^-$ state.} 
%\scalebox{0.5}{$^b$ Spin-parity of this state is unknown.}
\end{column}
\end{columns}
\caption{\label{table-mass-7} Experimental data and theoretical MCAS
results for $^7$Li and $^7$Be states (left table), and for $^7$He and $^7$B (right table).
All energies are in MeV and relate to scattering thresholds for 
nucleon+${}^6$He or nucleon+${}^6$Be. For states labeled by ``?$^a$'' spin-parity attributes are
unknown or uncertain.}
\end{table}

%%%%%%%%%%%%%%%%%%%%%%%%%%%%%%%%%%%%%%%%%%%%%%%%%%%%%%%%%%%%%%%%%%%%%%%%%%%%%%%%%%%%
 In Table~\ref{table-mass-7} are shown the results  obtained~\cite{Canton06a} 
 with that CC  nucleon-nucleus potential model.
 One CC potential model has been used for all four nuclides.
 The results for $^7$Li/$^7$Be (and for $^7$He/$^7$B) differ solely
 by the effect of the central Coulomb field. Instead, if we compare
 results for the pairs  $^7$Li/$^7$He and  $^7$Be/$^7$B, they differ
 solely for the different action of the OPP term.
 For the two mass-7 bound systems, Pauli blocking is assumed in
 the $0s_{1/2}$ shells and all the remaining shells are 
 considered allowed, while for the two unbound systems 
 a more complex OPP scenario is considered: 
 the $0s_{1/2}$ shells are blocked, the $p$ shells are hindered,
 and only the higher shells are completely allowed. This hindrance
 of the $p$ shells could reflect the anomalous interaction
 of the neutron with $^6$He, which is a neutron halo. A similar situation
 could occur in the mirror case of $^7$B, with the interaction of a proton 
 with a $^6$Be-type core. In our calculation, we have made the hypothesis
 of a Pauli hindrance in the p-shells defined by the following parameters:
 $\lambda(0p_{3/2}[0^+_1;2^+_1;2^+_2])=17.6 MeV$, 
 $\lambda(0p_{1/2}[0^+_1])=36.0 MeV$, 
 $\lambda(0p_{1/2}[2^+_1;2^+_2])=5.6 MeV$. The extended nature 
 of the even-even mass-6 subsystems, 
 either weakly-bound ($^6$He) or unbound ($^6$Be), is approximately 
 reflected also in the geometric size of the potential parameters 
 of Table~\ref{OMparams}, with rather extended radius, diffuseness, 
 and quadrupole deformation.
 
 \section{Conclusions}
 The Sturmian-based MCAS approach has been applied
 to coupled-channel problems at low energy, using phenomenological
 potentials with macroscopic, collective-type, couplings. 
 But the method is sufficiently flexible that it could be applied also
 in the presence of nonlocal potentials, such as those microscopically
 generated.
 Using simplified  collective-type couplings, we have applied 
 the approach to stable nuclei, as well as to weakly-bound and to unbound 
 (with respect to the nucleon emission threshold)
 light nuclei. In the few cases considered, interesting results 
 have been obtained, sometimes with rather good
 reproduction of bound spectra and scattering observables, and often with 
 predictions that could stimulate new experiments. In this approach,
 the highly nonlocal OPP term is crucial in order to include, macroscopically,
 the effects of Pauli exchanges. Finally, for weakly-bound or unbound light 
 systems, the concept of Pauli hindrance is suggested. 
 This implies that the nuclei have partially occupied $p$-shells or
     $p$-wave proto-shells wherein  the Pauli principle
     neither forbids nor allows occupancy. To some
     extent, it represents a suppression of access to phase-space.

%\begin{frame}
%\centerline{States for $^7$He and $^7$Li with the same nuclear potential}
%\centerline{Only the OPP term changes: 
%\only<1>{\alert{for $^7$Li}}\only<2>{\alert{for $^7$He}}}
%{\bf Pauli Forbidden $\lambda\simeq 1 GeV$}
%\begin{eqnarray*}
%\alert{0s_{1/2} + 0^+_1} & \alert{0s_{1/2} + 2^+_1} & \alert{0s_{1/2} +
%2^+_2}  \\
%\end{eqnarray*}
%\only<1>{\bf Pauli Allowed $\lambda\simeq 0 MeV$}
%\only<2>{\bf Pauli Hindered $\lambda\simeq 1\div 10 MeV$}
%\only<1>{\begin{eqnarray*}
%{0p_{3/2} + 0^+_1} & {0p_{3/2} + 2^+_1} & {1p_{3/2} +2^+_2} \\
%{0p_{1/2} + 0^+_1} & {0p_{1/2} + 2^+_1} & {0p_{1/2} +2^+_2}
%\end{eqnarray*}
%$\ $\\
%$\ $\\
%$\ $}
%\only<2>{\begin{eqnarray*}
%\alert{0p_{3/2} + 0^+_1} &\alert{0p_{3/2} + 2^+_1}&\alert{1p_{3/2} +2^+_2} \\
%\alert{0p_{1/2} + 0^+_1} &\alert{0p_{1/2} + 2^+_1}&\alert{0p_{1/2} + 2^+_2}\\
%\end{eqnarray*}
%$\lambda(0p_{3/2}[0^+_1;2^+_1;2^+_2])=17.6 MeV$, \\
%$\lambda(0p_{1/2}[0^+_1])=36.0 MeV$, \\
%$\lambda(0p_{1/2}[2^+_1;2^+_2])=5.6 MeV$
%}

%\end{frame}

\end{document}